\documentstyle[preprint,aps]{revtex}
\tightenlines

\begin{document}
\title{On the correlation energies for two interacting electrons in a parabolic
quantum dot}
\author{Omar Mustafa}
\address{Department of Physics, Eastern Mediterranean University\\
G. Magusa, North Cyprus, Mersin 10 - Turkey\\
email: omar.mustafa@emu.edu.tr\\
PACS; {73.63.Kv, 73.63.-b, 73.90.+f}}
\date{\today}
\maketitle

\begin{abstract}
{\small The correlation energies for two interacting electrons in a
parabolic quantum dot are studied via a pseudo-perturbation recipe. It is
shown that the central spike term, $(m^2-1/4)/r^2$, plays a distinctive role
in determining the spectral properties of the above problem. The study is
carried out for a wide range of the Coulomb coupling strength $\lambda$
relative to the confinement.}
\end{abstract}

\newpage

Advances in semiconductor technology have made it possible to fabricate
ultrasmall structures that confine electrons on a scale comparable to their
de Broglie wavelength. Structures that restrict the motion of electrons in
all directions are called quantum dots (QDs). The simplest QD consists of an
insulator, e.g. AlGaAs, and a semiconductor, e.g. GaAs. Between is a
potential difference that confines injected electrons to a thin layer at the
interface, in which electronic motion in the direction perpendicular to the 
{\em sandwich} is essentially {\em frozen out}. Effectively, QDs are
considered two-dimensional systems demonstrating typical quantum effects
such as discrete energy levels and interference. As a realistic and
computationally convenient approximation, a harmonic shape of the laterally
confined potential ( i.e., a two-dimensional oscillator effective in the
plane of the dot) is often used [1].

Several experimental [2-6] and theoretical [7-23] methods were invested for
the study of spectroscopic structure of interacting electrons in a harmonic
QD. For example, electron correlation has been investigated by the
many-particle Schr\"odinger equation [20], by perturbation [21], shifted 1/N
expansion [17], and by WKB treatments and exact numerical solutions [16]. QD
helium ground state in a magnetic field is obtained by a Hartree,
Hartree-Fock, and exact treatments [22]. Magnetic field dependance of
electron energies in QD is studied by the decoupled approximation [23],
screening of an ionic potential in QD [24], ${\cdots}$ etc.

In this paper we consider the simplest nontrivial problem of two interacting
electrons, with effective mass $m^{\ast }$, in a harmonic quantum dot. The
Hamiltonian of which is known to decouple into an exactly solvable
center-of-mass Hamiltonian\newline
\begin{equation}
H_{R}=\frac{P_{R}^{2}}{2M}+\frac{1}{2}M\omega _{o}^{2}~R^{2},
\end{equation}
\newline
and a non-exactly ( or, at best, conditionally-exactly) solvable Hamiltonian%
\newline
\begin{equation}
H_{r}=\frac{P_{r}^{2}}{2\mu }+\frac{1}{2}~\mu ~\omega _{o}^{2}~r^{2}+\frac{%
e^{2}}{\epsilon r}.
\end{equation}
\newline
Where, $\omega _{o}$ is the characteristic frequency of the parabolic
confinement, $\vec{R}=(\vec{r_{1}}+\vec{r_{2}})/2$, $\vec{P_{R}}=\vec{p_{1}}+%
\vec{p_{2}}$, $M=2m^{\ast }$, $\vec{r}=\vec{r_{1}}-\vec{r_{2}}$, $\vec{P_{r}}%
=(\vec{p_{1}}-\vec{p_{2}})/2$, and $\mu =m^{\ast }/2$. Moreover, when a
magnetic field $\vec{B}$ is applied perpendicular to the plane of the dot
one would simply amend the above Hamiltonians and replace $\omega _{o}$ by
the effective frequency $\tilde{\omega}=\sqrt{\omega _{o}^{2}+\omega
_{c}^{2}/4}$, where $\omega _{c}=eB/m^{\ast }c$ is the cyclotron frequency,
and the spin energy term $E_{s}=g^{\ast }\mu _{B}BS_{z}$ ( with $%
S_{z}=[1-(-1)^{m}]/2$, $g^{\ast }$ is the Land\'{e} factor, and $\mu _{B}$
is the Bohr magneton) could be added to the total energy of the dot.
However, upon the substitution $r=\sqrt{2}l_{o}q$, Schr\"{o}dinger equation
for the relative motion Hamiltonian (2) eventually reads \newline
\begin{equation}
\left[ -\frac{d^{2}}{dq^{2}}+\frac{m^{2}-1/4}{q^{2}}+q^{2}+\frac{\lambda }{q}%
\right] \Psi _{k,m}(q)=\Xi _{k,m}\Psi _{k,m}(q).
\end{equation}
\newline
Where $\Xi _{k,m}=2E_{k,m}/(\hbar \omega _{o})$, $E_{k,m}$ is the eigenvalue
of $H_{r}$ in (2), $\lambda =\sqrt{2}l_{o}/a^{\ast }$, $a^{\ast }=\hbar
^{2}\epsilon /(m^{\ast }e^{2})$ is the effective Bohr radius, and $l_{o}=%
\sqrt{\hbar /(m^{\ast }\omega _{o})}$ is the characteristic length of the
harmonic confinement.

It is well known, on the other hand, that results from perturbation theory
are limited to the case where $\lambda\ll1$ [10,21]. Moreover, the shifted
1/N expansion technique (SLNT) [17] and WKB treatments [16] lead to dubious
accuracies in connection with level ordering and energy crossings [16,25].
However, results from exactly solvable potentials can be used in
perturbation and pseudo-perturbation theories, or they can be combined with
numerical calculations. Nevertheless, in the simplest case, analytical and
semianalytical calculations can aid numerical studies in areas where
numerical techniques might not be safely controlled [26-31].

In this work we shall use a pseudo-perturbation theory (PSLET) to study the
spectroscopic structure of the relative motion Hamiltonian  of two
interacting electrons in a parabolic quantum dot, eq.(3). Where, in the
numerous methodical predecessors of a subset of papers [28-31], an
alternative possibility has been sought in the power-law asymptotic
expansions using some other {\em small} parameter. It has been noticed that
the presence of the central spike, e.g. $(m^2-1/4)/q^2$ in (3), just {\em %
copies} the effect of the centrifugal and/or centripetal force and
immediately inspires the use of {\em small} shifted inverse angular momentum
quantum number. An exhaustive description of the necessary formulae of PSLET
accompanied by the persuasive verifications of their numerical usefulness (
by immediate comparisons of its results with available {\em brute force}
numerical data) could be found in ref.s [28-31].

PSELT recipe starts with the augmentation of $m^2-1/4$, of the central spike
in (3), by $l_D(l_D+1)$. Where $l_D=l+(D-3)/2$, $D$ is the dimensionality, $%
l $ is angular momentum when $D=3$ and $l=|m|$ when $D=2$ (the
dimensionality under consideration). Then, we simply use $1/\bar{l}$ as a
pseudo-perturbation parameter, with $\bar{l}=l_D-\beta$ and $\beta$ is a
vital shift as it removes the poles that would emerge at $l=0$ for $D=3$.
Equation (3), with $V(q)=(q^2+\lambda/q)/2$, therefore reads\newline
\begin{equation}
\left\{-\frac{1}{2}\frac{d^{2}}{dq^{2}}+\frac{\bar{l}^{2}+(2\beta+1)\bar{l}
+\beta(\beta+1)}{2q^{2}}+\frac{\bar{l}^2}{Q}V(q) \right\} \Psi_{k,l} (q)=%
\bar{E}_{k,l}\Psi_{k,l}(q),
\end{equation}
\newline
where $\bar{E}_{k,l}=\Xi_{k,l}/2$ and Q is a constant that scales the
potential $V(q)$ at large - $l_D$ limit and is set equal to $\bar{l}^2$ at
the end of the calculations, for any specific choice of $l_D$ and nodal
zeros $k$. Next, we shift the origin of the coordinate system through $x=%
\bar{l}^{1/2}(q-q_{o})/q_{o}$, where $q_o$ is currently an arbitrary point
to be determined below. Expansions about this point, $x=0$ (i.e. $q=q_o$)
would lead to\newline
\begin{equation}
\left[-\frac{1}{2}\frac{d^{2}}{dx^{2}}+\frac{q_{o}^{2}}{\bar{l}} \tilde{V}%
(x(q))\right] \Psi_{k,l}(x) =\frac{q_{o}^2}{\bar{l}}\bar{E}%
_{k,l}\Psi_{k,l}(x),
\end{equation}
\newline
with\newline
\begin{equation}
\frac{q_o^2}{\bar{l}}\tilde{V}(x(q))= q_o^2\bar{l}\left[\frac{1}{2q_o^2}+%
\frac{V(q_o)}{Q}\right] +\bar{l}^{1/2}B_1 x+\sum^{\infty}_{n=0} v^{(n)}(x) 
\bar{l}^{-n/2},
\end{equation}
\newline
where\newline
\begin{equation}
v^{(0)}(x)=B_2 x^2 + \frac{2\beta+1}{2},
\end{equation}
\newline
\begin{eqnarray}
v^{(n)}(x)&=&B_{n+2}~ x^{n+2}+(-1)^n~ (2\beta+1)~ \frac{(n+1)}{2}~ x^n 
\nonumber \\
&+&(-1)^{n}~ \frac{\beta(\beta+1)}{2}~ (n-1)~ x^{(n-2)}~~;~~n \geq 1,
\end{eqnarray}
\newline
\begin{equation}
B_n=(-1)^n \frac{(n+1)}{2} +\left(\frac{d^n V(q_o)}{dq_o^n}\right)\frac{%
q_o^{n+2}}{n! Q}.
\end{equation}
\newline
It is then convenient to expand $\bar{E}_{k,l}$ as\newline
\begin{equation}
\bar{E}_{k,l}=\sum^{\infty}_{n=-2}E_{k,l}^{(n)}~\bar{l}^{-n}.
\end{equation}
\newline
Equation (5), along with (6-9), is evidently the one - dimensional
Schr\"odinger equation for a harmonic oscillator $\Omega^2 x^2/2$, with $%
\Omega^2=2B_2$, and the remaining terms in Eq.(5) are considered as an
infinite power series perturbation to the harmonic oscillator. One would
then imply that\newline
\begin{equation}
E_{k,l}^{(-2)}=\frac{1}{2q_o^2}+\frac{V(q_o)}{Q}
\end{equation}
\newline
\begin{equation}
E_{k,l}^{(-1)}=\frac{1}{q_o^2}\left[\frac{2\beta+1}{2} +(k +\frac{1}{2}%
)\Omega\right]
\end{equation}
\newline
Where $q_o$ is chosen to minimize $E_{k,l}^{(-2)}$, i. e.\newline
\begin{equation}
\frac{dE_{k,l}^{(-2)}}{dq_o}=0~~~~ and~~~~\frac{d^2 E_{k,l}^{(-2)}}{dq_o^2}%
>0.
\end{equation}
\newline
Equation (13) in turn gives, with $\bar{l}=\sqrt{Q}$,\newline
\begin{equation}
l_D-\beta=\sqrt{q_{o}^{3}V^{^{\prime}}(q_{o})}.
\end{equation}
\newline
The shifting parameter $\beta$ is determined by choosing $\bar{l}%
E_{k,l}^{(-1)}$=0. Hence\newline
\begin{equation}
\beta=-\left[\frac{1}{2}+(k+\frac{1}{2})\Omega\right],~~ \Omega=\sqrt{3+%
\frac{q_o V^{^{\prime\prime}}(q_o)}{V^{^{\prime}}(q_o)}}
\end{equation}
\newline
where primes of $V(q_o)$ denote derivatives with respect to $q_o$. Then
equation (5) reduces to\newline
\begin{equation}
\left[-\frac{1}{2}\frac{d^2}{dx^2} + \sum^{\infty}_{n=0} v^{(n)} \bar{l}%
^{-n/2}\right]\Psi_{k,l} (x)= \left[\sum^{\infty}_{n=1} q_o^2
E_{k,l}^{(n-1)} \bar{l}^{-n} \right] \Psi_{k,l}(x).
\end{equation}
\newline
Setting the wave functions with any number of nodes $k$ as \newline
\begin{equation}
\Psi_{k,l}(x(q)) = F_{k,l}(x)~ exp(U_{k,l}(x)),
\end{equation}
\newline
equation (16) readily transforms into the following Riccati equation:\newline
\begin{eqnarray}
&&F_{k,l}(x)\left[-\frac{1}{2}\left(
U_{k,l}^{^{\prime\prime}}(x)+U_{k,l}^{^{\prime}}(x)
U_{k,l}^{^{\prime}}(x)\right) +\sum^{\infty}_{n=0} v^{(n)}(x) \bar{l}^{-n/2}
\right.  \nonumber \\
&&\left. -\sum^{\infty}_{n=1} q_o^2 E_{k,l}^{(n-1)} \bar{l}^{-n} \right]
-F_{k,l}^{^{\prime}}(x)U_{k,l}^{^{\prime}}(x)-\frac{1}{2}F_{k,l}^{^{\prime%
\prime}}(x)=0,
\end{eqnarray}
\newline
where the primes denote derivatives with respect to $x$. It is evident that
this equation admits solution of the form \newline
\begin{equation}
U_{k,l}^{^{\prime}}(x)=\sum^{\infty}_{n=0} U_{k}^{(n)}(x)~~\bar{l}^{-n/2}
+\sum^{\infty}_{n=0} G_{k}^{(n)}(x)~~\bar{l}^{-(n+1)/2},
\end{equation}
\newline
\begin{equation}
F_{k,l}(x)=x^k +\sum^{\infty}_{n=0}\sum^{k-1}_{p=0} a_{p,k}^{(n)}~~x^p~~\bar{%
l}^{-n/2},
\end{equation}
\newline
\begin{equation}
U_{k}^{(n)}(x)=\sum^{n+1}_{m=0} D_{m,n,k}~~x^{2m-1} ~~~~;~~~D_{0,n,k}=0,
\end{equation}
\newline
\begin{equation}
G_{k}^{(n)}(x)=\sum^{n+1}_{m=0} C_{m,n,k}~~x^{2m}.
\end{equation}
\newline
Substituting equations (19) - (22) into equation (16) implies\newline
\begin{eqnarray}
&&F_{k,l}(x)\left[-\frac{1}{2}\sum^{\infty}_{n=0}\left(U_{k}^{(n)^{^{%
\prime}}} \bar{l}^{-n/2} + G_{k}^{(n)^{^{\prime}}} \bar{l}^{-(n+1)/2}\right)
\right.  \nonumber \\
&-&\left.\frac{1}{2} \sum^{\infty}_{n=0} \sum^{n}_{m=0} \left(
U_{k}^{(m)}U_{k}^{(n-m)} \bar{l}^{-n/2} +G_{k}^{(m)}G_{k}^{(n-m)} \bar{l}%
^{-(n+2)/2} \right. \right.  \nonumber \\
&+&\left.\left.2 U_{k}^{(m)}G_{k}^{(n-m)} \bar{l}^{-(n+1)/2}\right)
+\sum^{\infty}_{n=0}v^{(n)} \bar{l}^{-n/2} -\sum^{\infty}_{n=1} q_o^2
E_{k,l}^{(n-1)} \bar{l}^{-n}\right]  \nonumber \\
&-&F_{k,l}^{^{\prime}}(x)\left[\sum^{\infty}_{n=0}\left(U_{k}^{(n)}\bar{l}%
^{-n/2} + G_{k}^{(n)} \bar{l}^{-(n+1)/2}\right)\right]-\frac{1}{2}%
F_{k,l}^{^{\prime\prime}}(x) =0
\end{eqnarray}
\newline
The solution of equation (22) follows from the uniqueness of power series
representation. Therefore, for a given $k$ we equate the coefficients of the
same powers of $\bar{l}$ and $x$, respectively.

Although the energy series, equation (23), could appear divergent, or, at
best, asymptotic for small $\bar{l}$, one can still calculate the
eigenenergies to a very good accuracy by forming the sophisticated Pad\'e
approximation\newline
\begin{equation}
P_{N}^{M}(1/\bar{l})=(P_0+P_1/\bar{l}+\cdots+P_M/\bar{l}^M)/ (1+q_1/\bar{l}%
+\cdots+q_N/\bar{l}^N)
\end{equation}
\newline
to the energy series (23). The energy series is calculated up to $%
E_{k,l}^{(11)}/\bar{l}^{11}$ by 
\begin{equation}
E_{k,l}=\bar{l}^{2}E_{k,l}^{(-2)}+E_{k,l}^{(0)}+\cdots +E_{k,l}^{(11)}/\bar{l%
}^{11}+O(1/\bar{l}^{12}),
\end{equation}
\newline
and with the $P_{5}^{5}(1/\bar{l})$ Pad\'e approximant it becomes\newline
\begin{equation}
E_{k,l}[10,9]=\bar{l}^{2}E_{k,l}^{(-2)}+P_{5}^{5}(1/\bar{l}).
\end{equation}
\newline

Following the above procedure, PSLET results are compared, in table 1, with
the exact numerical ones ( obtained by direct numerical integrations, DNI)
[16] for $\lambda=1$ and $\lambda=10$. To avoid exhaustive numbers of tables
we do not list Garcia-Castelan et. al's results [16] from WKB, WKB
single-parabola (WKB-SP), and WKB double-parabola (WKB-DP). In contrast with
the WKB, WKB-SP, WKB-DP [16], and SLNT [13] results the comparison between
PSLET and DNI results implies excellent agreement.

In order to make remediable analysis on the effect of $\lambda $, hence of
the characteristic length $l_{o}$ ($\lambda \sim \l _{o}$), we list ( in
tables II-IV) PSLET results for $k=0,1,2$ and $\lambda =0,1,2,4,6,8,10,12$
at different values of $|m|$. They are also plotted in figure 1.

Figure 1 (along with tables II-IV) shows that the degeneracies associated
with the harmonic oscillator confinement at $\lambda=0$ are only partially
lifted as $\lambda $ increases from zero ( of course, such degeneracies
would completely be lifted when a magnetic field is applied perpendicular to
the plane of the dot). It also shows that the equidistance form of the
energy levels at $\lambda =0$ changes in the following manners; (i) for a
given $k$, the spacing between two successive $|m|$ states decreases as $%
\lambda $ increases, and increases as $|m|$ increases for a given $\lambda $%
, whilst (ii) for a given $\lambda ,$ the spacing increases as the nodal
quantum number $k$ increases. One should nevertheless notice that (iii)
s-states (with $m=0$) shift up more rapidly than states with $|m|\geq 1$,
and for $|m|\geq 1$ states with lower $|m|$ shift up faster than states with
higher $|m|$  as $\lambda $ increases from zero.

The above mentioned features (i)-(iii), in fact, build up  the sought after
scenario for the change in level ordering, that manifests energy crossings
and spin-singlet ($S_{z}=0$) spin-triplet ($S_{z}=1$) oscillations, and
inspires the vital role of the central spike term in (3). More specifically,
the {\em twofold} nature of the central spike term in the effective
potential, of eq.(3),\newline
\begin{equation}
V_{eff}(q)=\frac{m^{2}-1/4}{q^{2}}+q^{2}+\frac{\lambda }{q}
\end{equation}
\newline
explains the energy crossings as follows; (a) for $m=0$ it represents an
attractive core that strengthens the confinement $q^{2}$, whereas (b) for $%
|m|\geq 1$ it represents a repulsive core which renders, along with the
Coulomb repulsion, the potential less potent. This is why, for a given k,
the energy of a lower $|m|$ state increases much faster ( more rapidly for $%
m=0$) than that of a higher $|m|$, as $\lambda $ increases, and catches up
with it ( hence energy crossings and singlet-triplet spin oscillations
occur, or, at most, energy levels clustering is manifested). On the physical
sides, the two electrons are farther apart for higher $|m|$. Moreover, for a
given $k$ energy crossings are not feasible between the corresponding states
with different $|m|$. Whereas, states with a given $k$ and $|m|$ cross with
states at lower $k$ and higher $|m|$. Therefore, the lowest three states
(0,0), (0,1), and (0,2) never cross any other state ( i.e., they can never
be depressed into a lower $k$-state).

The effect of correlation, between two interacting electrons in a harmonic
QD, is therefore clear in the full energy spectrum for $\lambda >0$ with all
($k$,$|m|$)-states for the relative motion as shown in figure 1 and
documented in tables I-IV. However, it should be noted that the level
ordering reported by Garcia-Castelan et. al [15] is now changed, namely for
the (0,4) and (1,1) states. Moreover, the (1,4) and (2,1) states seem to
change order as $\lambda $ increases form 12.

To sum up, we have used a pseudo-perturbation recipe (PSLET) to study the
characteristic length effect on the correlation energies for two interacting
electrons in a parabolic QD. We have proved PSLET persuasive numerical
reliability in comparison with direct numerical integration method ( in
table 1) for $\lambda $=1, and 10. Next, we have obtained the correlation
energies for $\lambda =2,4,6,8,12$ and $k=0,1,2$, and documented ( through
figure I) that the level ordering reported by Garcia-Castelan et. al [15] is
not absolute but continually bound to change as $\lambda $ increases from
zero. Finally, the almost forgotten {\em twofold} effect of the central
spike term, in the effective two-dimensional potential (27), is now
clarified to inherit a major responsibility for energy crossings.

\newpage 
\begin{table}[tbp]
\caption{Comparison of PSLET energies (in $\hbar\protect\omega_o/2$ units)
and the exact ones from direct numerical integration [16] for $\protect%
\lambda$=1 and 10.}
\begin{center}
\vspace{1cm} 
\begin{tabular}{|cccccc|}
\hline\hline
& $\lambda$=1 &  &  & $\lambda$=10 &  \\ \hline
$(k,|m|)$ & Exact & PSLET & $(k,|m|)$ & Exact & PSLET \\ \hline
(1,7) & 20.3587 & 20.3587 & (1,7) & 23.5040 & 23.5040 \\ 
(0,9) & 20.3280 & 20.3280 & (0,9) & 23.2188 & 23.2188 \\ 
(3,2) & 18.5351 & 18.5351 & (3,2) & 23.0339 & 23.0339 \\ 
(2,4) & 18.4388 & 18.4388 & (2,4) & 22.2217 & 22.2217 \\ 
(1,6) & 18.3843 & 18.3843 & (3,1) & 21.8721 & 21.8715 \\ 
(0,8) & 18.3472 & 18.3472 & (1,6) & 21.7355 & 21.7355 \\ 
(3,1) & 16.6498 & 16.6498 & (0,8) & 21.3954 & 21.3954 \\ 
(2,3) & 16.4895 & 16.4895 & (3,0) & 21.3513 & 21.3140 \\ 
(1,5) & 16.4163 & 16.4163 & (2,3) & 20.6504 & 20.6504 \\ 
(0,7) & 16.3701 & 16.3701 & (1,5) & 20.0186 & 20.0186 \\ 
(3,0) & 14.9850 & 14.9881 & (0,7) & 19.6037 & 19.6037 \\ 
(2,2) & 14.5646 & 14.5646 & (2,2) & 19.2438 & 19.2438 \\ 
(1,4) & 14.4579 & 14.4579 & (1,4) & 18.3753 & 18.3753 \\ 
(0,6) & 14.3983 & 14.3983 & (2,1) & 18.1420 & 18.1420 \\ 
(2,1) & 12.6961 & 12.6961 & (0,6) & 17.8543 & 17.8543 \\ 
(1,3) & 12.5154 & 12.5154 & (2,0) & 17.6671 & 17.6660 \\ 
(0,5) & 12.4340 & 12.4340 & (1,3) & 16.8431 & 16.8431 \\ 
(2,0) & 11.0848 & 11.0883 & (0,5) & 16.1628 & 16.1628 \\ 
(1,2) & 10.6024 & 10.6024 & (1,2) & 15.4916 & 15.4916 \\ 
(0,4) & 10.4814 & 10.4814 & (0,4) & 14.5546 & 14.5547 \\ 
(1,1) & 8.7594 & 8.7594 & (1,1) & 14.4622 & 14.4622 \\ 
(0,3) & 8.5485 & 8.5485 & (1,0) & 14.0379 & 14.0381 \\ 
(1,0) & 7.2340 & 7.2362 & (0,3) & 13.0720 & 13.0720 \\ 
(0,2) & 6.6538 & 6.6538 & (0,2) & 11.7903 & 11.7903 \\ 
(0,1) & 4.8553 & 4.8553 & (0,1) & 10.8495 & 10.8496 \\ 
(0,0) & 3.4952 & 3.4968 & (0,0) & 10.4816 & 10.4816 \\ \hline\hline
\end{tabular}
\end{center}
\end{table}
\newpage 
\begin{table}[tbp]
\caption{ PSLET correlation energies (in $\hbar\protect\omega_o/2$ units)
for $k$=0, $|m|$=0,1,2,3,4,5 and $\protect\lambda$=0,1,2,4,6,8,10,12}
\begin{center}
\vspace{1cm} 
\begin{tabular}{|ccccc|}
\hline\hline
$|m|$ & $\lambda$=0 & $\lambda$=1 & $\lambda$=2 & $\lambda$=4 \\ \hline
0 & 2 & 3.4968 & 4.6391 & 6.4428 \\ 
1 & 4 & 4.8553 & 5.6557 & 7.1251 \\ 
2 & 6 & 6.6538 & 7.2872 & 8.4994 \\ 
3 & 8 & 8.5485 & 9.0864 & 10.1331 \\ 
4 & 10 & 10.4814 & 10.9564 & 11.8885 \\ 
5 & 12 & 12.4340 & 12.8638 & 13.7112 \\ \hline
& $\lambda$=6 & $\lambda$=8 & $\lambda$=10 & $\lambda$=12 \\ \hline
0 & 7.9373 & 9.2644 & 10.4816 & 11.6184 \\ 
1 & 8.4599 & 9.6938 & 10.8496 & 11.9425 \\ 
2 & 9.6480 & 10.7425 & 11.7903 & 12.7975 \\ 
3 & 11.1440 & 12.1226 & 13.0720 & 13.9947 \\ 
4 & 12.7978 & 13.6861 & 14.5547 & 15.4049 \\ 
5 & 14.5429 & 15.3599 & 16.1628 & 16.9525 \\ \hline\hline
\end{tabular}
\end{center}
\end{table}

\newpage 
\begin{table}[tbp]
\caption{ Same as table 2 for $k$=1.}
\begin{center}
\vspace{1cm} 
\begin{tabular}{|ccccc|}
\hline\hline
$|m|$ & $\lambda$=0 & $\lambda$=1 & $\lambda$=2 & $\lambda$=4 \\ \hline
0 & 6 & 7.2362 & 8.2945 & 10.0462 \\ 
1 & 8 & 8.7594 & 9.4879 & 10.8608 \\ 
2 & 10 & 10.6024 & 11.1913 & 12.3314 \\ 
3 & 12 & 12.5154 & 13.0233 & 14.0173 \\ 
4 & 14 & 14.4579 & 14.9110 & 15.8031 \\ \hline
& $\lambda$=6 & $\lambda$=8 & $\lambda$=10 & $\lambda$=12 \\ \hline
0 & 11.5189 & 12.8317 & 14.0381 & 15.1665 \\ 
1 & 12.1368 & 13.3327 & 14.4622 & 15.5362 \\ 
2 & 13.4252 & 14.4772 & 15.4916 & 16.4721 \\ 
3 & 14.9840 & 15.9254 & 16.8431 & 17.7388 \\ 
4 & 16.6773 & 17.5343 & 18.3753 & 19.2009 \\ \hline\hline
\end{tabular}
\end{center}
\end{table}
\newpage 
\begin{table}[tbp]
\caption{ Same as table 2 for $k$=2.}
\begin{center}
\vspace{1cm} 
\begin{tabular}{|ccccc|}
\hline\hline
$|m|$ & $\lambda$=0 & $\lambda$=1 & $\lambda$=2 & $\lambda$=4 \\ \hline
0 & 10 & 11.0883 & 12.0757 & 13.7327 \\ 
1 & 12 & 12.6961 & 13.3720 & 14.6650 \\ 
2 & 14 & 14.5646 & 15.1195 & 16.2015 \\ \hline
& $\lambda$=6 & $\lambda$=8 & $\lambda$=10 & $\lambda$=12 \\ \hline
0 & 15.1801 & 16.4736 & 17.6660 & 18.7833 \\ 
1 & 15.8856 & 17.0418 & 18.1420 & 19.1937 \\ 
2 & 17.2480 & 18.2613 & 19.2438 & 20.1978 \\ \hline\hline
\end{tabular}
\end{center}
\end{table}
\newpage

\begin{center}
{\bf {\Large Figures captions} }
\end{center}

{\bf Fig.1:} PSLET correlation energies for two interacting electrons in a
harmonic quantum dot vs $\lambda =\sqrt{2}l_{o}/a^{\ast }$ (the ratio of the
oscillator length $l_{o}$ and the effective Bohr radius $a^{\ast }$ . The
energies are normalized with the oscillator energy $\hbar \omega _{o}/2$.
The full lines represents states with $k=0$, dashed lines for $k=1$, and
dashed dotted lines for $k=2$.

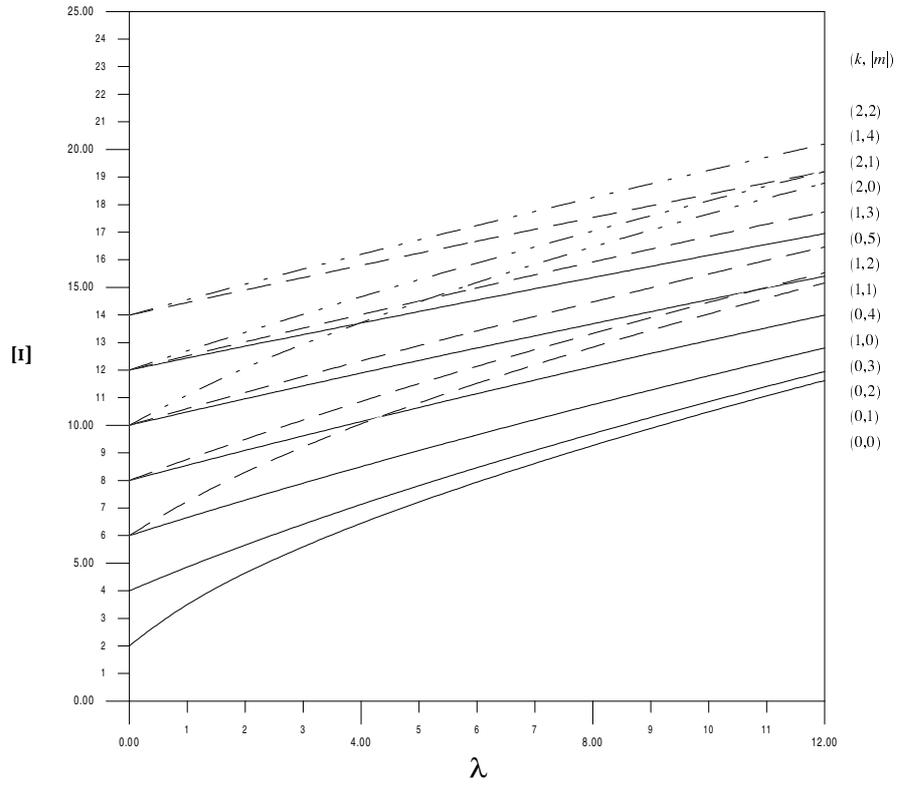
\end{document}